\documentclass[12pt]{article}
\usepackage{graphicx}
\usepackage{amssymb}
\usepackage{amsmath}
\usepackage{cite}
\usepackage{graphicx}
\usepackage{float}
\numberwithin{equation}{section}

\def\BR{{\mathbb R}}

\def\clq{{\mathcal Q}}

\def\clf{{\mathcal F}}

\newtheorem{Pa}{Paper}[section]

\newtheorem{Cy}[Pa]{{\bf Corollary}}
\newtheorem{Rk}[Pa]{{\bf Remark}}

\newtheorem{Ee}[Pa]{{\bf Example}}
\newtheorem{Dn}[Pa]{{\bf Definition}}
\newtheorem{Nn}[Pa]{{\bf Notations}}
\newtheorem{Pn}[Pa]{{\bf Proposition}}

\begin{document}

\title{Electron-photon interaction: Feynman diagrams and contact points}

\author{ Lev Sakhnovich}

\date{}

\maketitle


\noindent \emph{99 Cove ave. Milford, CT 06461, USA, }\\
\noindent E-mail: lsakhnovich@gmail.com\\

\noindent {\bf Keywords:} Feynman diagram, direct problem, inverse problem, contact point, colored graph.

\date{}

\maketitle


 \noindent\textbf{MSC (2020):} 81T18,   81Q30.

\date{}
\begin{abstract}
In this note, we formulate the notions of the direct and inverse problems and contact points
for the Feynman diagrams.
For the electron-photon interaction case, the solutions of   these  direct and inverse problems
are presented. The interrelations between
Feynman diagrams and the classical  colored graphs are discussed.
\end{abstract}

\section{Introduction.}
{\bf 1.}  In \cite[section 4]{Sak1}  a first approximation $a_{1}(L,q)$
of the scattering function 
\begin{align}& d(L,q,\varepsilon)=1+\varepsilon\big(a_{1}(L,q)\big)+o(\varepsilon) \quad (L>0, \, q\in \BR^4) \label{1.1}\end{align}
is considered for the ultraviolet case. (Here, $\BR$ stands for the real axis.)
See also, for instance,  \cite{AB, Dy, Fey, Opp, Sak}) on this QED topic. 
The corresponding scattering operator is the operator of the multiplication by the scattering function in the space
of functions of $q$. The first approximation $a_{1}(L,q)$ may be written down as an integral over the four dimensional sphere with radius $L$ (in spherical coordinates):
\begin{equation}
a_{1}(L,q)=-i\int_{0}^{L}\int_{0}^{2\pi}\int_{0}^{\pi}\int_{0}^{\pi}\clf(p,q)r^{3}\big(\sin^{2}\phi_{1}\big)\sin\phi_2d\phi_1d\phi_2d\phi_3dr,\label{1.2}
\end{equation}
where $\clf(p, q)$ is a rational matrix function, $p=[p_1,p_2,p_3,p_4],\,\, q=[q_1,q_2,q_3,q_4]$ $(p,q\in \BR^4)$ and
\begin{align}& p_1=r\cos\phi_1,\quad p_2=r\sin\phi_1\cos\phi_2, \label{4.3-}
\\ &
p_3=r\sin\phi_1\sin\phi_2\cos\phi_3, \quad
p_4=r\sin\phi_1\sin\phi_2\sin\phi_3 .\label{1.3}\end{align}
In the classical case, it is assumed that the limit of  $d(L,q,\varepsilon)$ $(L\to \infty)$ exists: $d(q,\varepsilon)=\lim_{L\to \infty}d(L,q,\varepsilon)$. The nonclassical case is much more complicated. In 
 the case of the ultraviolet divergence,  the limit of $a_1(L,q)$ (for $L\to \infty$) does not exist. 
 
 R. Feinman used his diagrams in order to recover the first approximation $a_1(q,L)$. In \cite{Sak1}, we construct the scattering operator
 from the first  approximation. Brought together, these approaches may be used  in order to construct  the scattering operator from the
 Feynman diagrams (direct problem). In a wider context, Feynman diagrams  are closely connected with important direct and inverse
 spectral problems.  
 
 More precisely, Feynman diagrams include external lines (i.e.,  inverse data, which is an analog of the spectral data) and internal lines
(i.e. direct data, which is an analog of the ``potential" in the case of differential equations). Clearly, the scattering operator is a
spectral data. Direct problems are the problems to recover the spectral data or other inverse data (from the direct data) and  inverse problems are the problems
to recover  direct data from the inverse data. Some of these problems are treated in our present work.

We note that in our papers \cite{Sak, Sak2, Sak1}  (as well as here) there are no differential equations, which could be considered as
direct data.
 
{\bf  2.} In this paper, instead of \eqref{1.2} we study the case of the $4m$-fold integral:
\begin{equation} ia_1(L,Q)=J=\int{\clf(P_1,Q_1,P_2,Q_2,...,P_m,Q_m)}d^4{P_1}d^4{P_2}...d^4{P_m}\label{1.5},\end{equation}
where $P_k,Q_k \in \BR^4, \quad Q=[Q_1,Q_2, \ldots , Q_m]\in \BR^{4m}$  and the function $\break \clf(P_1,Q_1,P_2,Q_2,...,P_m,Q_m)$ is rational.
Integral J may be  written down as an integral over the $4m$-fold sphere with radius L (in spherical coordinates).
If $m=1$, we set $P_1=p$, $Q_1=q$ and \eqref{1.5} takes the form \eqref{1.2}.

Within the framework of Feynman theory,  we will consider  direct and inverse problems
and will introduce  the so called contact points which play an important role in the formulation of these
problems.
It is characteristic that contact points belong  to the external and  internal data, both. 
\section{Feynman diagrams and contact points}
In the seminal    paper \cite{Fey}  by R. Feynman, specific physical problems 
 are associated with Feynman diagrams (see also \cite{AB} and the references therein).  Recall that Feynman diagrams 
 consist of  directed external lines, undirected internal  lines, and vortices.  The external lines correspond to the observable outgoing or  incoming particles and begin or end, respectively, at some vertex of the diagram. 
 The internal lines correspond to the virtual particles, which are not directly observed. An internal line connects two vortices
 and the external line is connected with only one vertex. 
 \begin{Dn}\label {Dn2.1}
An irreducible Feynman diagram is a diagram that cannot be split into two
disconnected parts by removing certain internal line.\end{Dn} 
 
  In the case of the electron-photon interactions, there are external and internal electron and photon lines (both types of lines).
 We introduce the following notations for the corresponding Feynman diagram.
 \begin{Nn} \label{Nn} The number of the external electron  lines is denoted by $N_e$.  The number of the external photon  lines is 
 denoted by $N_p$. The number of the internal electron  lines is denoted by $F_e$.  The number of the internal photon  lines is 
 denoted by $F_p$. The number of the vertices of the inner part of the diagram is  $n$.
\end{Nn}
Next, we formulate an important proposition \cite{AB}.
 \begin{Pn}\label{Pn2.3} Let an integral $J$  of the form \eqref{1.5} correspond to the electron-photon interaction $($and the corresponding Feynman diagram$)$.
 Then, the power of the numerator of  $ \clf(P_1,Q_1,P_2,Q_2,...,P_m,Q_m)$ equals $F_e$ and the power of its denominator
 equals 
  \begin{equation}F=F_e+F_p .\label{2.1}\end{equation}
 \end{Pn}
The {\it power counter} $K$ may be introduced by the formula
\begin{equation}K=2F-(F_e+4m). \label{2.2}\end{equation}
The numbers $F_e,F_p,N_e,N_p, K, n$
characterize the topological properties of the Feynman diagram. 

Further we assume that the Feynman diagram is irreducible and the following main condition is fulfilled (see \cite{AB} for the corresponding 
important cases in QED).

{\bf Main condition.} {\it  At  each vertex of the Feynman diagram, two electron lines and one photon line are meeting. Of those lines  two are
internal and one is external.}
  
Thus, the vertices at which internal electron and photon lines make contact coincide with the vertices, where  electron lines are either incoming or outgoing.  {\it These vertices are called  contact points and their number is denoted by $\clq$.}

 Since the contact points are the points
at which internal electron and photon lines make contact, the contact points and their number $\clq$ belong to the internal data.
Since the contact points are the points, where  electron lines are either incoming or outgoing,  the contact points and their number $\clq$ belong to the external data.

We see that no more than one external electron line may correspond to any vortex. 
If this external line exists, such a vortex is a contact point and internal electron and photon lines meet at that vortex as well. If  a vortex is not
a contact point, an external photon line and two internal electron lines meet there.
Therefore, recalling Notation \ref{Nn} we have 
\begin{equation}N_e=\clq,\quad  F_e=\big(n+(n-\clq)\big)/2=n-\clq/2, \quad F_p=\clq/2,\quad N_p=n-\clq.\label{2.3}\end{equation}
It follows from \eqref{2.3}  that
\begin{equation} 2F_e+N_e=2n ,\quad 2F_p+N_p=n, \quad F_e+F_p=n ,\quad N_e+N_p=n.\label{2.5}\end{equation}
The first and third equalities in  \eqref{2.3} imply  the corollary below.
\begin{Cy}\label{Corollary 2.2}The numbers $N_e$ and $\clq$ are even.\end{Cy}
Relations \eqref{2.1}--\eqref{2.3} yield
\begin{equation} K=2F_p+F_e-4m=n+\clq/2 - 4m=(3/2)\clq+n-\clq-4m.\label{2.6}\end{equation}
Hence, using again \eqref{2.3}, we obtain the next assertion.
\begin{Cy}\label{Corllary {2.3}} The counter number $K$ may be represented as$:$
\begin{equation} K=(3/2)N_e+N_p -4m .\label{2.7}\end{equation}
\end{Cy}
As already mentioned in the introduction, external data is observable and internal data is unobservable.
{\it Using  \eqref{2.3} and \eqref{2.5}, we solve the inverse problem to  recover the unobservable internal data
 ($\clq,n,F_e,F_p$) from the  observable external data ($N_e,N_p$)}:
\begin{equation} \clq=N_e,  \quad n=N_e+N_p, \quad  F_p=N_e/2,  \quad F_e=N_p+N_e/2 .\label{2.9}\end{equation}
\begin{Rk}\label{Remark 2.6}  Recall that a classical inverse problem is to recover a differential equation from the spectral data.
We do  not consider this problem here.\end{Rk}
Next, we solve a {\it direct problem} to recover the quantities $n,\clq, N_e,N_p$  from the unobservable quantities
$F_e$ and $F_p$. Namely, using again \eqref{2.3} and \eqref{2.5} we derive
\begin{equation}\clq=2F_p,  \quad n=F_e+F_p, \quad N_e=2F_p, \quad N_p=F_e-F_p.\label {2.10}\end{equation}
\begin{Rk} \label{RkDI} We already mentioned that the numbers $\clq$ and $n$ belong in our case to direct and inverse
data, both. It follows from \eqref{2.9} and \eqref{2.10} that the complete direct data is recovered from the inverse data
$N_e$ and $N_p$, and  the complete inverse data is recovered from the direct data $F_e$ and $F_p$. 
\end{Rk}
\begin{Rk} \label{RkF}
According to Furry's theorem \cite{Fur},  Feynman diagrams have physical sense
only for the even numbers $n$.
\end{Rk}
\section{The counter number \\ and divergence condition}
{\bf 1.} According to Corollary \ref{Corollary 2.2}, Remark \ref{RkF} and the last equality in \eqref{2.5}, we have
the following proposition.
\begin{Pn}\label{Propostion 3.1}The numbers $N_e, N_p$ and $n$ are even. \end{Pn}
Next, let us study the integral \eqref{1.5}. Recall that the counter number $K$ is given by \eqref{2.2} or, equivalently, \eqref{2.7}.
The statement on the convergence/divergence of the integral \eqref{1.5} below follows from \cite{AP}.

1) \emph{The integral \eqref{1.5}  is divergent if and only if $K\leq 0$. }
Hence,  the integral \eqref{1.5} is divergent when 
\begin{equation}  (3/2)N_e+N_p \leq 4m.\label{3.1}\end{equation}

Now, we set $m=1$ in \eqref{1.5}.  According to Furry's theorem (see Remark~\ref{RkF}) and \eqref{3.1},
we  have statement 2) below.

2) {\it The inequality $K\leq 0$ and the corresponding divergence  may hold only in the cases $n=0,\, 2, \, 4$.}

Finally, we formulate statement 3)  (see  \cite{AP}).

 3) {\it Only four cases of divergence  exist for the interaction studied in this note with $m=1$  $($see below$)$.}\\
Case 1. (logarithmic divergence): \\
$n=4, \quad N_e=0,\quad N_p=4, \quad K=0, \quad F_e=4,\quad F_p=0,\quad F=4$. \\
Case 2. (linear divergence): \\
$n=2, \quad N_e=2,\quad N_p=0, \quad K=-1,\quad F_e=F_p=1,\quad F=2$.\\
Case 3. (quadratical divergence): \\
$n=2, \quad N_e=0, \quad N_p=2, \quad K=-2, \quad F_e=2,\quad F_p=0,\quad F=2$. \\
Case 4. (vacuum‑to‑vacuum transition, trivial case): \\
$n=0, \quad N_e=N_p=0, \quad K=-4,\quad F_e=F_p=F=0$.

\begin{Ee}\label{Example 3.1}Let us consider the integral $($see \cite[§ 47.1]{AB}$):$
\begin{equation}J_{\mu}=\int_{\Omega}\frac{p_{\mu}d^{4}p}{(p^2-2pq+\ell(q))^2}=i\pi^{2}q_{\mu}\left(\ln\frac{L^2}{\ell(q)-q^2}-\frac{3}{2}\right),
\label{3.3} \end{equation}
where $\Omega$ is a 4-fold sphere of radius $L$; $\, \ell(q)>q^2$;  $\, \, p^2$, $q^2$ and $pq$ are scalar products,  and $\mu=1,2,3$ or $4$.
\end{Ee}
According to Proposition \ref{Pn2.3}, we have $F_e=1$ and  $F=2$, that is, $F_p=1$.  The solution of the direct problem for this case
(see \eqref{2.10}) yields that 
$$\clq=n=2, \quad N_e=2, \quad N_p=0.$$
The corresponding Feynman diagram has the following form:

\begin{figure}[h!]
\includegraphics[width=10cm]{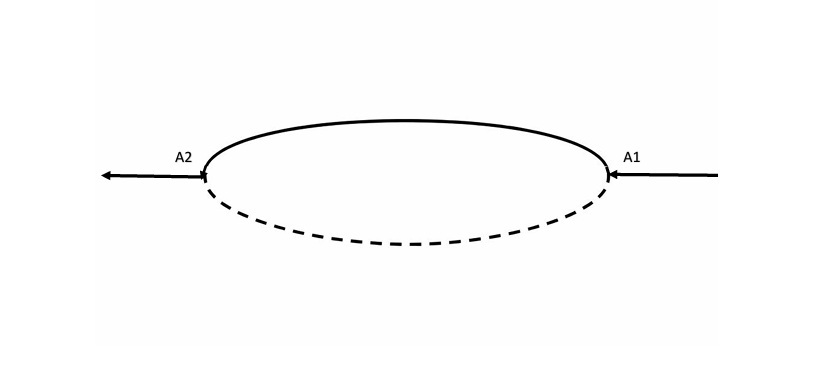}
\centering
\caption{Case 2.}
\label{fig:fig1}
\end{figure}
Here, $ A_1$ and $A_2$ are the vertices (and contact points), the straight lines  are electron lines and  the dashed line is the photon line.

From the said above, one may see that the integral in \eqref{3.3}  corresponds to the Case 2.
However, the divergence in \eqref{3.3} is not linear because the coefficient by $L$  on the right hand side equals zero. (Thus, the linear term is omitted.) 
Clearly, the divergence in \eqref{3.3} is logarithmic.  

If one considers {\it the inverse problem} for the case 2,  one starts with the {\it inverse data}: $N_e=2, \quad N_p=0$ and,
using \eqref{2.9}, obtains the direct data:
$$\clq=n=2, \quad F_e=F_p=1.$$

\begin{Ee}\label{Example 3,4} Consider  the Feynman diagram below $($i.e.,  Figure 2$):$

\begin{figure}[h!]
\includegraphics[width=8cm]{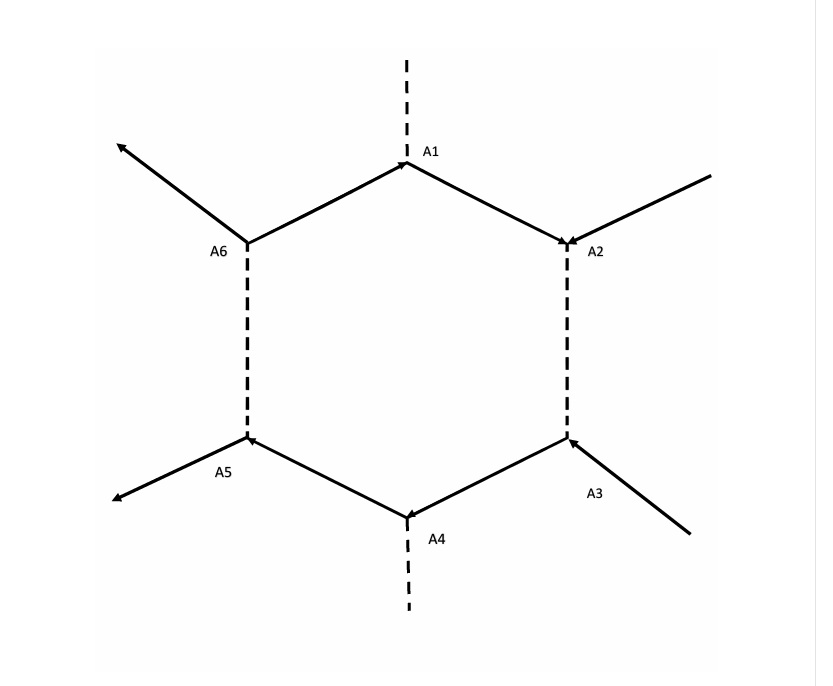}
\centering
\caption{The case of a converging integral}
\label{fig:fig1}
\end{figure}
Here, $n=6$ and the number $\clq$ of the contact points  $A_2,\,A_3,\, A_5,\,A_6$
equals $4$. Moreover, we see that
\begin{equation}    F_e=4,\quad F_p=2,    \quad     N_e=4,\quad N_p=2,           \label{z1}\end{equation}
and $m=1$ for the corresponding Feynman integral.
\end{Ee} 
Clearly, the direct data $\{\clq, \, n , \, F_e, \, F_p\}$ may be recovered from $N_e$ and $N_p$ via \eqref{2.9}.
The inverse data  $\{\clq, \, n , \, N_e, \, N_p\}$ may be recovered from $F_e$ and $F_p$ via \eqref{2.10}.

The Main condition from the previous section is satisfied for this Feynman diagram, and we have $m=1$. Thus, taking into account
statement 3) from this section and the fact that $n=6$, we derive that the corresponding Feynman integral
converges.

{\bf 2.}  {\it Feynman diagrams and graphs.}
Let us consider the Feynman diagram as a graph. The number $m$ of independent loops
(i.e., independent closed circuits or, equivalently, independent closed paths formed by the internal lines) of a Feynman diagram  is called
the first Betti number or   cyclomatic number  in the graph theory.  Recall that $F=F_e+F_p$ is the number of the internal lines.
Therefore, graph theory yields the following formula for $m$:  
\begin{equation}m=F-n+C,\label{2.14},\end{equation} 
$C$ is  the number of the connected components of the graph.
Using the third equality in \eqref{2.5} and \eqref{2.14}, we obtain
\begin{equation}m=C \label{2.15} \end{equation}
in our case.

In the case of our Feynman diagrams, we have for kinds of lines: photon lines (external and internal) and electron lines (external and internal).
Thus, these Feynman diagrams are modifications of the $4$-colored  graphs. However, they differ from the classical colored case in two ways:

1. Feynman diagrams have important external lines and the classical  colored  graphs have not.

2.  In Feynman diagrams, two lines of the same color may meet at some vertices, and
in the classical  colored  graphs  they cannot.

\vspace{0.3em}

The {\it main problem} to solve for the colored graphs is to  find the smallest number of colors
for the given graph. {\it In Feynman theory the main problem is the problem to recover
Feinman diagram from the given external data, and  to construct the scattering operator
for this diagram.}

\end{document}